# An Automated, Cost-Effective Optical System for Accelerated Antimicrobial Susceptibility Testing (AST) using Deep Learning

Calvin Brown[1], Derek Tseng[1], Paige M. K. Larkin[2], Susan Realegeno[2], Leanne Mortimer[2], Arjun Subramonian[3], Dino Di Carlo[4,5,6], Omai B. Garner[2], Aydogan Ozcan[1,4,5,7,*]

[1]Department of Electrical and Computer Engineering, [2]Department of Pathology and Laboratory Medicine, [3]Department of Computer Science, [4]Department of Bioengineering, [5]California NanoSystems Institute (CNSI), [6]Jonsson Comprehensive Cancer Center, University of California, Los Angeles, CA 90095, and [7]Department of Surgery, David Geffen School of Medicine, University of California, Los Angeles, CA 90095

* Corresponding author: ozcan@ucla.edu

**ABSTRACT:** Antimicrobial susceptibility testing (AST) is a standard clinical procedure used to quantify antimicrobial resistance (AMR). Currently, the gold standard method requires incubation for 18–24 h and subsequent inspection for growth by a trained medical technologist. We demonstrate an automated, cost-effective optical system that delivers early AST results, minimizing incubation time and eliminating human errors, while remaining compatible with standard phenotypic assay workflow. The system is composed of cost-effective components and eliminates the need for optomechanical scanning. A neural network processes the captured optical intensity information from an array of fiber optic cables to determine whether bacterial growth has occurred in each well of a 96-well microplate. When the system was blindly tested on isolates from 33 patients with *Staphylococcus aureus* infections, 95.03% of all the wells containing growth were correctly identified using our neural network, with an average of 5.72 h of incubation time required to identify growth. 90% of all wells (growth and no-growth) were correctly classified after 7 h, and 95% after 10.5 h. Our deep learning-based optical system met the FDA-defined criteria for essential and categorical agreements for all 14 antibiotics tested after an average of 6.13 h and 6.98 h, respectively. Furthermore, our system met the FDA criteria for major and very major error rates for 11 of 12 possible drugs after an average of 4.02 h, and 9 of 13 possible drugs after an average of 9.39 h, respectively. This system could enable faster, inexpensive, automated AST, especially in resource-limited settings, helping to mitigate the rise of global AMR.

## Introduction

Antimicrobial resistance (AMR) is estimated to cause over 700,000 deaths annually, with 2.8 million cases and 35,000 deaths in the United States alone.[1,2] By 2050, the number of deaths due to AMR is projected to reach as many as 10 million per year.[1] A host of factors are contributing to the global rise in AMR, such as over-prescription and abuse of antibiotics (e.g. for viral infections),[3,4] use of medically important antibiotics in agriculture for e.g. promotion of growth in livestock[5] and prevention of disease in citrus trees,[6] as well as economic and regulatory barriers to the development of new drugs.[7]

One of the most crucial tools to treat patients infected with resistant bacteria, as well as to stem the tide of global AMR, is antimicrobial susceptibility testing (AST). AST is a laboratory procedure used to determine which antibiotics will work most effectively against a given patient's bacterial infection. The gold standard method is broth microdilution (BMD), in which isolated patient bacteria are inoculated in growth medium along with a candidate antibiotic and incubated for at least 18–24 h. BMD is usually performed in a 96-well microplate, with a different antibiotic/concentration combination in each well. Neighboring wells contain successive two-fold dilutions of the same drug. After incubation, each well is inspected visually by a trained medical technologist to determine whether growth has occurred, as indicated by the presence of turbidity in the well. The minimum inhibitory concentration (MIC) for a given antibiotic is defined as the lowest concentration of the drug that successfully prevents bacterial growth. The MIC is used to determine the categorical susceptibility of the bacteria to the drug (susceptible, intermediate, or resistant) based on concentration cutoffs published by the Clinical & Laboratory Standards Institute (CLSI).[8]

The lengthy incubation time (18–24 h or more) puts patients at risk because in the interim they may be prescribed powerful broad-spectrum antibiotics or antibiotics against which the organism is resistant. The need for a trained expert to manually read the plate strains laboratory resources and inevitably introduces human error/variability. Automated AST systems such as the bioMérieux Vitek 2[9] enable readings much earlier during incubation for certain bacteria and drugs, but these systems are relatively bulky, expensive (due to e.g. optomechanical scanning components and illumination sources), and often require the use of proprietary plates and drug panels, limiting their utility especially in resource-limited settings, where AMR is expected to take the largest toll.[10]

Numerous alternative approaches have been investigated to address the shortcomings of conventional AST. The decreasing cost of whole-genome sequencing (WGS) has made it a potentially-viable option, and it has been shown to agree with BMD for certain bacteria-drug combinations.[11-13] However, the cost remains prohibitive for most labs, even in developed countries, and there is a lack of standardization for AST protocols.[14] In addition, unless all the resistance mechanisms in question for a given sample are linked to genes with well-characterized effects—such as the *mecA* gene in methicillin-resistant *Staphylococcus aureus* (MRSA)[15]— WGS-based AST will provide an incomplete resistance profile, limiting its applications, especially for *emerging forms of resistance*.[16-18] Due to its ability to enable rapid, low-cost diagnostics using small sample volumes, microfluidic technology has also been investigated for AST. By confining bacteria to microscale channels or droplets, the incubation time required to identify the impact of antibiotics on bacterial growth can be shortened considerably.[19-27] MICs can be determined straight from positive cultures (without the additional overnight isolation step) in the case of urine samples, but not for more complex samples such as blood or sputum.[26,27] Additionally, these microfluidic approaches generally require new specialized consumables and a scanning microscopy system to monitor the sample during incubation, limiting their viability in resource-limited settings.

Pure microscopy-based approaches have also been demonstrated for AST.[28-34] Commercially available systems such as the Pheno (Accelerate Diagnostics)[35] and the oCelloScope (BioSense Solutions)[36] have developed a more compact form factor compared to benchtop microscopes, but still require expensive objective lenses and optomechanical scanning components to read a 96-well plate. These systems also depend on knowledge of specific organism morphologies and growth characteristics, limiting their use to certain types of bacteria. As an alternative, lensfree microscopy[37-39] eliminates the need for objective lenses, thus reducing costs and mitigating the spatial/focal drift these components can cause during time lapse imaging. Lensfree microscopy has been shown to detect bacteria over a wide field of view,[40,41] but has not yet been demonstrated for AST. Previously, we also demonstrated a smartphone reader for AST plates *after* the incubation period to determine turbidity results.[42,43] These earlier works did not capture time lapse images of the samples and therefore were aimed to provide end-point readings, after the standard incubation period (e.g., 18–24 h).

In this work, we demonstrate an automated, cost-effective optical system for the *early detection* and *quantification* of resistance in AST using deep learning. The device can be placed directly inside a standard benchtop incubator and automatically monitor growth in all 96 wells of a standard microplate during incubation. The plate is periodically illuminated by red, green, and blue LEDs, and the transmitted light is relayed by an array of plastic optical fibers beneath the plate to two Raspberry Pi cameras for imaging. A neural network uses the intensity information from the images to classify each well as either turbid or non-turbid over time. This system eliminates the need to wait 18–24 h or more, offers significant time savings and does *not* rely on a trained medical technologist for readings as is necessary for conventional AST, while also being compact and cost-effective compared to commercially available automated AST systems.

Our system was blindly tested on 33 unique clinical *Staphylococcus aureus* isolates, using a panel containing varying concentrations of 14 antibiotics. 95.03% of all wells containing growth were correctly identified, with an average of 5.72 h of incubation required to identify growth. 90% of all wells were correctly classified after 7 h, and 95% after 10.5 h. The system met the FDA-defined criteria[44] for essential and categorical agreement for all 14 drugs tested after an average of 6.13 h and 6.98 h, respectively. The system met FDA criteria for major and very major error rates for 11 of 12 possible drugs after an average of 4.02 h, and 9 of 13 possible drugs after an average of 9.39 h, respectively. For each one of the drugs that did not meet the FDA criteria, only a single major or very major error was made. Some of the major and very major errors may also be due to human errors in the ground truth reading. With additional training and testing samples, the FDA criteria could potentially be met for all drugs. This system could enable inexpensive, high-throughput AST in resource-limited settings, helping treat infected patients while curbing the rise of drug-resistant bacteria.

## Results
### Imaging System Design
The AST system (Figure 1a,b) is composed of cost-effective components: LEDs, plastic optical fibers, singlet lenses, Raspberry Pi computers and camera modules, and 3D printed housing. It can be placed inside any standard laboratory incubator (Figure 1c) and has a slot for the insertion of a standard 96-well microplate loaded with bacterial isolates, growth medium, and candidate antibiotics at various concentrations (Figure 1d). Two adjacent 8x8 RGB LED arrays illuminate the entire plate from above, with one LED centered over each well. A plastic diffuser beneath the LEDs ensures spatial uniformity of illumination over the wells, and the brightness of each LED is controlled by pulse width modulation to compensate for the fact that wells near the center of the plate receive more light (due to neighboring LEDs) than those at the edge. Wells containing bacterial growth scatter the incident illumination, while the wells with no growth allow the light to pass through mostly unobstructed. Below each well, a bundle of 21 plastic optical fibers (Figure 1e) relays the transmitted light to one of two larger bundles, which are each imaged by the combination of a singlet lens and a CMOS camera connected to a Raspberry Pi computer. A sample image is shown in Figure 1f. Images are periodically captured over the course of an 18 h incubation, and examples of fiber intensity changes over time for wells with and without turbidity are shown in Figures 1g-i.

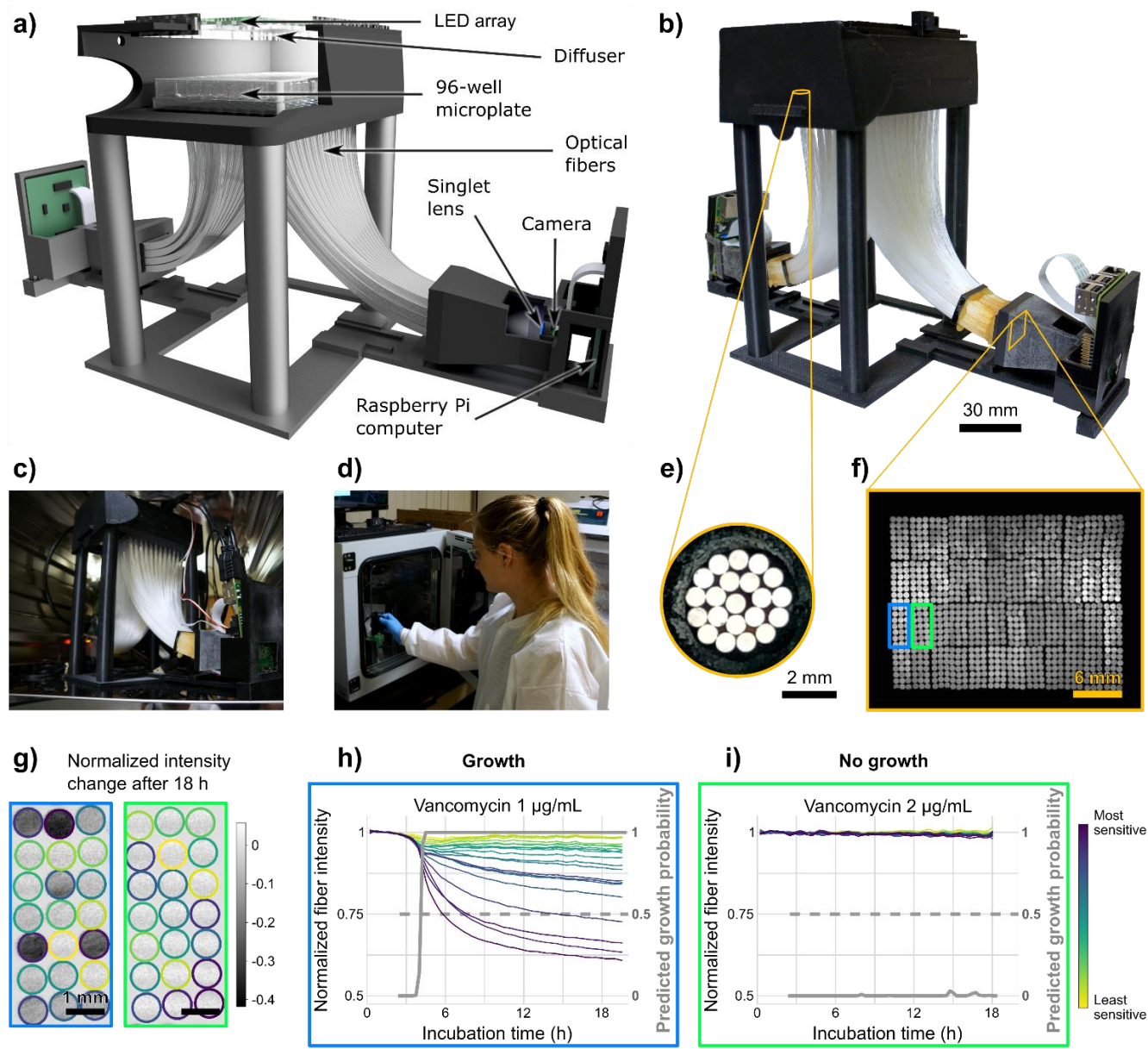

Figure 1. (a) Schematic and (b) photo of the device. (c) Device inside an incubator. (d) 96-well plate being loaded. (e) Close-up of 21 fibers under one well. (f) Image of fibers captured by the system. (g) Normalized fiber intensity change after 18 h incubation for two neighboring wells. Fiber intensities and neural network predicted probability of turbidity for (h) a turbid well and (i) a non-turbid well. The colormap corresponds to the random arrangement of fibers in each well. The predicted growth probability on the right axis corresponds to the gray curve in each plot, which is the output of the neural network as a function of the incubation time.

In addition to capturing images, the two Raspberry Pis synchronously control the schedule of the illumination and image capture during incubation (Figure 2). Every five minutes, the LEDs are turned on and an image is captured, enabling temporal sampling of potential growth while ensuring that the bacteria are not exposed to phototoxic levels of light. The illumination cycles through the three LED colors, so that the time between images of the same color is 15 minutes. A quality control strain of *S. aureus* was run repeatedly in the system to ensure the MICs were in the expected ranges,[45] indicating bacterial growth is not hampered by the periodic illumination (see Table S1).

The fiber array functions to demagnify the plate area, enabling imaging of all 96 wells *without* any optomechanical scanning components, while maintaining a compact form factor.[42,43] In this case, the fibers provide a demagnification factor of ~7, while capturing spatial information within each well, which is especially important for wells showing weak or atypical growth. The number of fibers per well (21) and the focal length of the singlet lenses (50 mm) were selected to maximize the amount of information captured per well. To address future manufacturability concerns, the locations of each of the 21 fibers within the wells were not manually recorded and tracked during assembly. Instead, the rough position of the fibers within each well (and thus, the information content) was empirically determined *post hoc* from the training data (further detailed below).

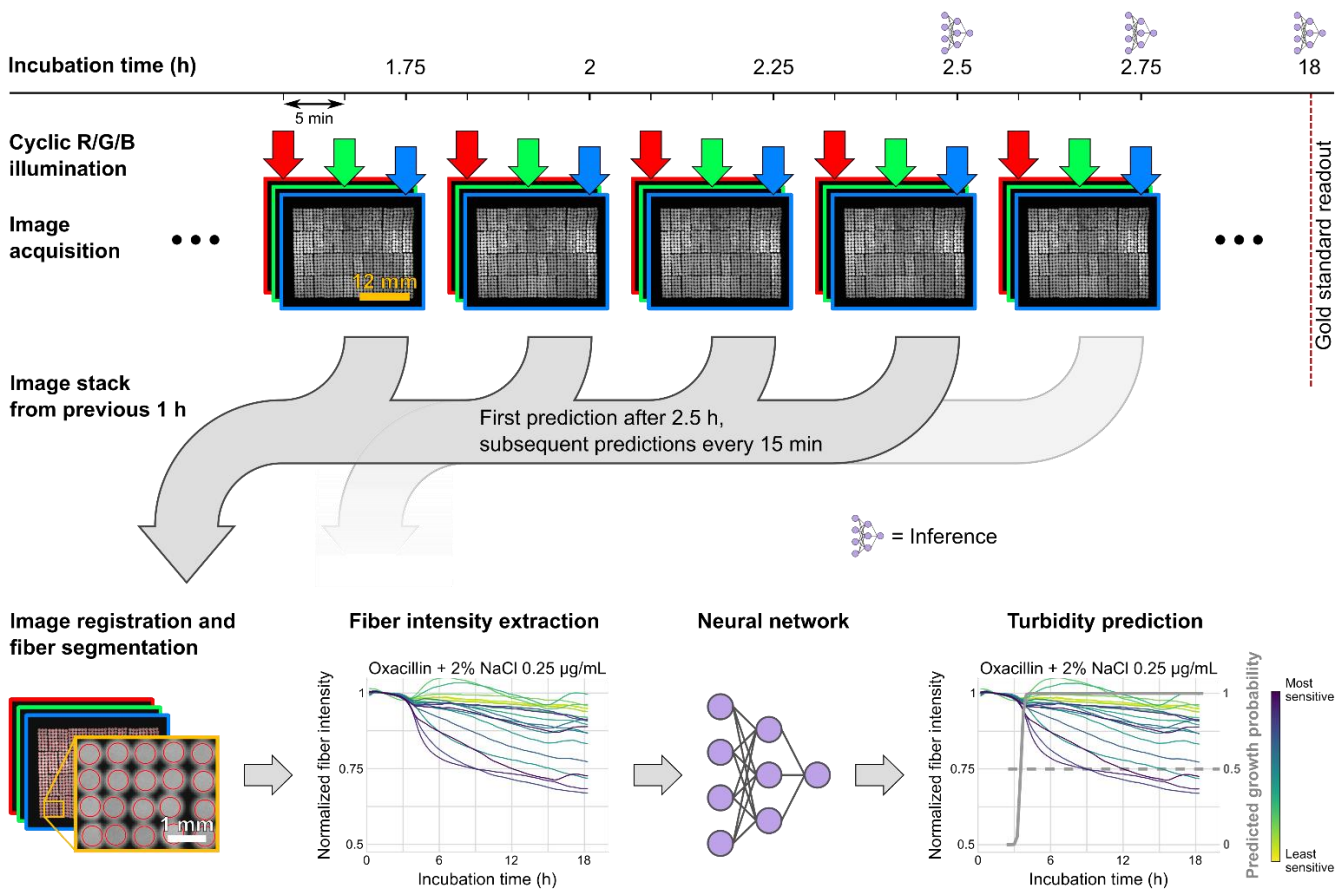

Figure 2. Image processing pipeline. Images are captured every 5 min under either red, green, or blue illumination. Images are registered and fiber intensities are extracted. A neural network uses the fiber intensities from the previous 1 h to predict the probability that each well is turbid at the current time point.

The entire device measures 175 × 450 × 192 mm and the cost of the components (including all optics, electronics, and 3D printing) is under $500, which would drop significantly at higher manufacturing volume. Our system easily integrates with a typical clinical workflow, using any standard laboratory incubator and standard 96-well microplates. To operate our system, a user simply inserts a plate (Figure 1d), then starts the image acquisition program on one of the Raspberry Pi computers. Our system was successfully operated by five different clinical laboratory personnel.

**Image Processing and Neural Network Design**

The data processing pipeline is shown in Figure 2. For each image, only the pixels corresponding to the illumination color are used (either red, green, or blue from the Bayer color filter array). For each plate, all subsequent images are aligned with the first image of the corresponding color using intensity-based registration, to account for any drift that may occur due to e.g. plate insertion, thermal effects, structural vibrations, etc. The first image of each color is also used to locate each fiber using the circular Hough transform, constrained by prior knowledge of the fiber grid layout. Using these fiber locations, the intensity of each fiber in each image is determined by averaging over a circular mask with a radius of 8 pixels, significantly smaller than the radius of the fiber to avoid any edge effects. Each temporal fiber intensity is denoised with a 30-minute moving averaging filter and normalized to its average value over the first 10 images during incubation (2.5 h), during which time detectable turbidity is not expected to develop. These preprocessing steps mitigate the effect of fiber intensity variation due to illumination, fiber polishing defects, off-axis effects, etc.

A turbidity prediction is made for each well after each image (every 5 min) starting after 2.5 h of incubation. The normalized fiber intensities for all 3 illumination colors over the previous one hour are fed into a neural network that outputs a predicted probability of turbidity in the well at the current time. This is referred to as the window slicing method in the time series classification literature.[46] *Any value above 0.5 is interpreted as turbid, while values below 0.5 are interpreted as non-turbid*. Examples of fiber intensity plots and blind testing network predictions are shown in Figures 3, S1, and S2. The turbidity classifications are then used to determine the MIC and susceptibility for each drug based on established clinical cutoffs.[8] The neural network comprises 4 fully connected hidden layers of 128 neurons each, and a binary classification output layer (Figure S3). Batch normalization and dropout (probability 0.5) were used after each hidden layer to accelerate training and limit overfitting, respectively. The network was trained with the Adam optimizer at a starting learning rate of 1e-3, which was decreased after the validation loss failed to improve for 20 epochs. In total, the network has 83,073 trainable parameters. Note that the network does not employ any spatial convolutional layers because the information contained in the fiber bundle images is not shift-invariant: each fiber corresponds to a fixed region of the plate. This is the reason the extracted fiber intensities—as opposed to images— are used as the input to the network. The network does not receive any prior information about the well, drug, or drug concentration; it makes predictions in a "well-blind" manner, which prevents it from overfitting to the specifics of the plates that were used in the experiments.

Training neural networks via supervised learning requires ground truth labels for every training sample. Because a ground truth reading can only be performed via visual inspection by the trained medical technologist *after* incubation, ground truth labels were only available for the final time point of each patient plate (~18 h). Labels for the training plates at every other image time point during incubation were created manually by inspecting the fiber intensity plots for each well, such as those in Figures 3, S1, and S2. While these labels do not constitute a ground truth, they are the best available proxy and were used to train the network to identify turbidity effectively at an earlier time point within the incubation phase. Additionally, in certain instances where the ground truth label after 18 h disagreed with the manual label, the label was changed for network training. This type of data cleaning is acceptable (and common) for training/validation data, especially when access to ground truth is not available, but certainly must not be (and was not) employed on blind testing data as it could bias results.

As mentioned previously, the 21-fiber bundles under each well were assembled without precise control of the mapping of each fiber from the well to the image to make the device easier to potentially manufacture in large quantities. In the imaged fiber bundles (Figure 1f), the fibers are grouped by well, but the fibers within each well are randomly arranged. For each well, a "fiber order" was determined empirically, using the amount by which the fiber intensities dropped in the presence of turbidity, averaged over the entire training set. The fibers that show the largest drop can be assumed to be near the center of the well, where growth will ultimately be strongest by the end of incubation, while the fibers that show the smallest drop can be assumed to be near the edges of the well, where even strong growth will have a limited effect on transmission (due to the settling that occurs in the round-bottomed wells for Gram-positive bacteria). The resulting fiber ordering was used for training and blind testing of the network and can be seen in all fiber intensity plots. This fiber ordering ensures that the network learns a general, robust model of turbidity over all wells, instead of overfitting to the individual characteristics of the fibers of each well.

While the total number of training samples, each representing a single time point from a single well from a single patient plate, was large (263,019), the number of clinical isolates from which the training data was gathered was smaller (51). The variability among isolates accounted for much of the diversity of the dataset, both because each isolate had a unique resistance profile and because each plate was incubated on a different day by a set of rotating technicians. To ensure that the learned network model was robust to this isolate-to-isolate variability, we employed nine-fold cross-validation by randomly splitting the 51 clinical isolate plates into nine subsets and training nine models, where each model was trained on eight of the subsets and validated on the ninth. This process was repeated 50 times and the best model for each subset was selected to form a final composite "panel" of nine neural networks. This cross-validation/composite method (a type of model bagging)[47,48] was employed to ensure that no single isolate plate exerted undue influence on the final model, as could have been the case if only a single validation set were used. Additionally, training in nine folds ensured each model was trained on a large number of dates (e.g., 45), which we found to improve performance on the validation data (Figure S4). Training many models was feasible because the number of layers and weights is small compared to many state-of-the-art image classification networks, which contain hundreds of millions of parameters. All training of the models was performed on a desktop computer in TensorFlow without a graphics processing unit (GPU), and in the future, due to improvements in computational power, could even be performed on the Raspberry Pi.

**Clinical Testing Results**

All experiments were performed at the UCLA Clinical Microbiology Laboratory by clinical staff, using 96-well microplates containing a Gram-positive antibiotic panel. *Staphylococcus aureus* isolates were prepared to a 0.5 McFarland standard in sterile water and then diluted in Mueller Hinton Broth. The diluted suspension was pipetted into all 96 wells and the plate was inserted into the AST system inside an incubator. 96-well microplates contained antibiotics in powder form pre-loaded into each well of the plate. Bacteria were pipetted into all 96 wells along with growth medium and inserted into the AST system inside an incubator. The plate was removed and ground truth reading was performed after 18–19 h. Initial experiments were performed on 47 plates, each with a quality control strain of *S. aureus* with a known resistance profile to ensure the system was functioning properly and the bacteria were not experiencing phototoxicity (Table S1). The MICs obtained from these control runs showed that the antibiotic linezolid used in the plates did not perform as expected and failed the quality control assessment, so its wells were excluded from the study. Any wells containing antimicrobials that do not have interpretive criteria for *S. aureus* and are not routinely used for clinical management (e.g. ceftriaxone) were also excluded from the study. Next, 51 plates, each containing a *S. aureus* isolate from a unique patient, were used to generate training and validation data for the neural network. 33 additional patient plates were used for blind testing data. The blind testing plates were read after 18–19 h by *two trained technologists* and wells with discrepant readings between the technologists were not used for testing the system (Table S2). A single technologist was used to determine the turbidity ground truth for the training/validation data.

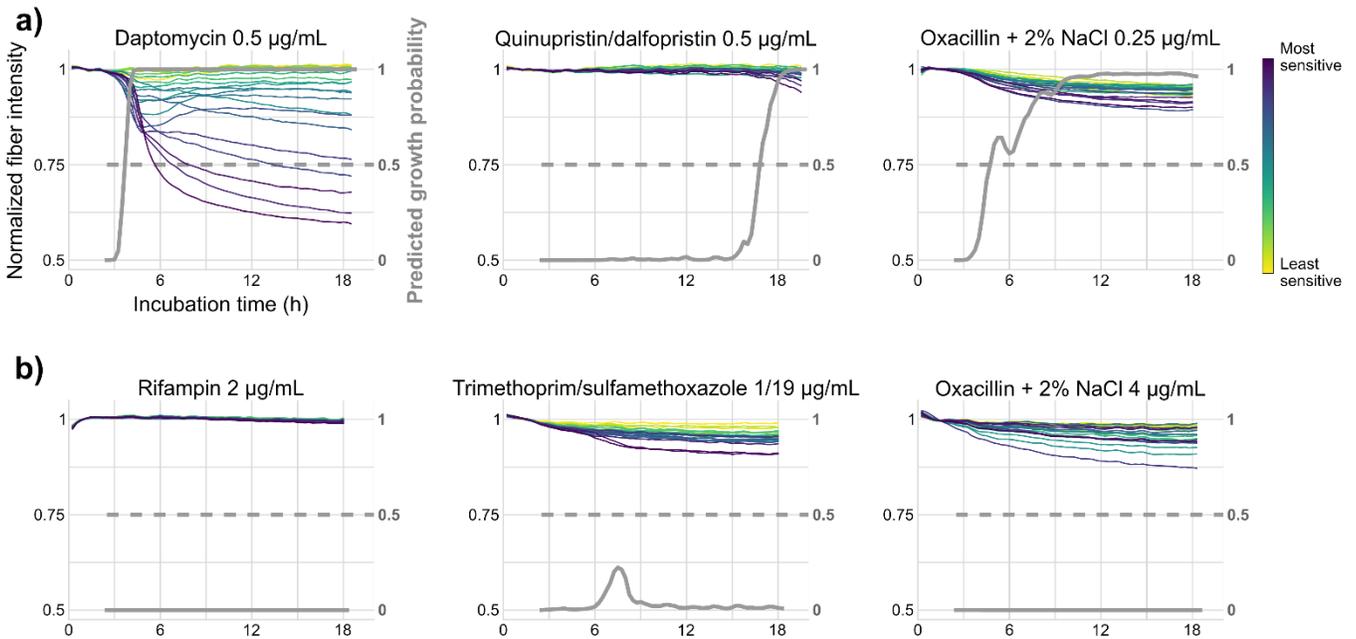

Figure 3. Fiber intensities and the panel of neural networks' predicted probability of turbidity on blind testing patient isolates of *Staphylococcus aureus* for ground truth (a) turbid and (b) non-turbid wells.

Examples of fiber intensities and neural network turbidity predictions for blindly tested patient plates are shown in Figures 3, S1, and S2. The average incubation time required to obtain a correct turbid prediction for each drug is shown in Figure 4a. 95.03% of all turbid wells were correctly identified by the network, with the average turbid well requiring just 5.72h of incubation to detect. The system detected turbidity for oxacillin in an average of 4.5 h, while it required 9 h on average for trimethoprim/sulfamethoxazole (Bactrim). Figure 4b shows the turbidity detection accuracy over time for all drugs. 90% of all wells were correctly classified after 7 h and 95% after 10.5 h. The MIC and susceptibility predictions for each drug were compared to the FDA-defined criteria for automated AST systems, namely essential agreement (EA), categorical agreement (CA), major error (maj) rate, and very major error (vmj) rate.[44] EA is the percentage of patients for which a drug's predicted MIC is within plus or minus one two-fold dilution of the ground truth. CA is the percentage of patients for which the predicted susceptibility category (susceptible/intermediate/resistant) matches the ground truth. Maj rate is the percentage of all susceptible infections misclassified as resistant (i.e. false positive) and vmj rate is the percentage of all resistant infections misclassified as susceptible (i.e. false negative). The FDA requires automated AST systems to demonstrate EA and CA greater than 90%, and maj rate and vmj rate of no more than 3%.

Figure 5 shows the blind testing results for EA, CA, maj rate, and vmj rate for each of the 14 drugs over the course of incubation. The legend in each plot indicates the number of valid samples in the denominator of the calculation for each drug. For EA/CA this is the number of blind testing patient plates for which there was agreement between the two human readers (out of a possible 33), and for maj/vmj rate it is the number of susceptible and resistant patient infections for which the two readers agreed, respectively. EA and CA surpassed the FDA limit of 90% for all 14 drugs well before the end of incubation (18–19 h). Note that EA/CA began near 0% at the beginning of incubation for drugs against which growth/resistance was common such as daptomycin, whereas EA/CA began higher for drugs against which growth/resistance was rare, such as rifampin.

Maj rate remained below the FDA limit of 3% for 11 of 12 possible drugs, and the drug for which it exceeded 3% (trimethoprim/sulfamethoxazole or Bactrim) was due to a single major error. It was not possible to calculate maj rate for two drugs (gatifloxacin and erythromycin) because re-sistance to these drugs was not observed for any clinical isolates. The maj rate for each of the other 12 drugs is plotted, but those that never move above 0% obscure one another.

Vmj rate dropped below the FDA-permitted maximum of 3% before the end of incubation for 9 of 13 possible drugs. Again, the four drugs for which the system did not meet the

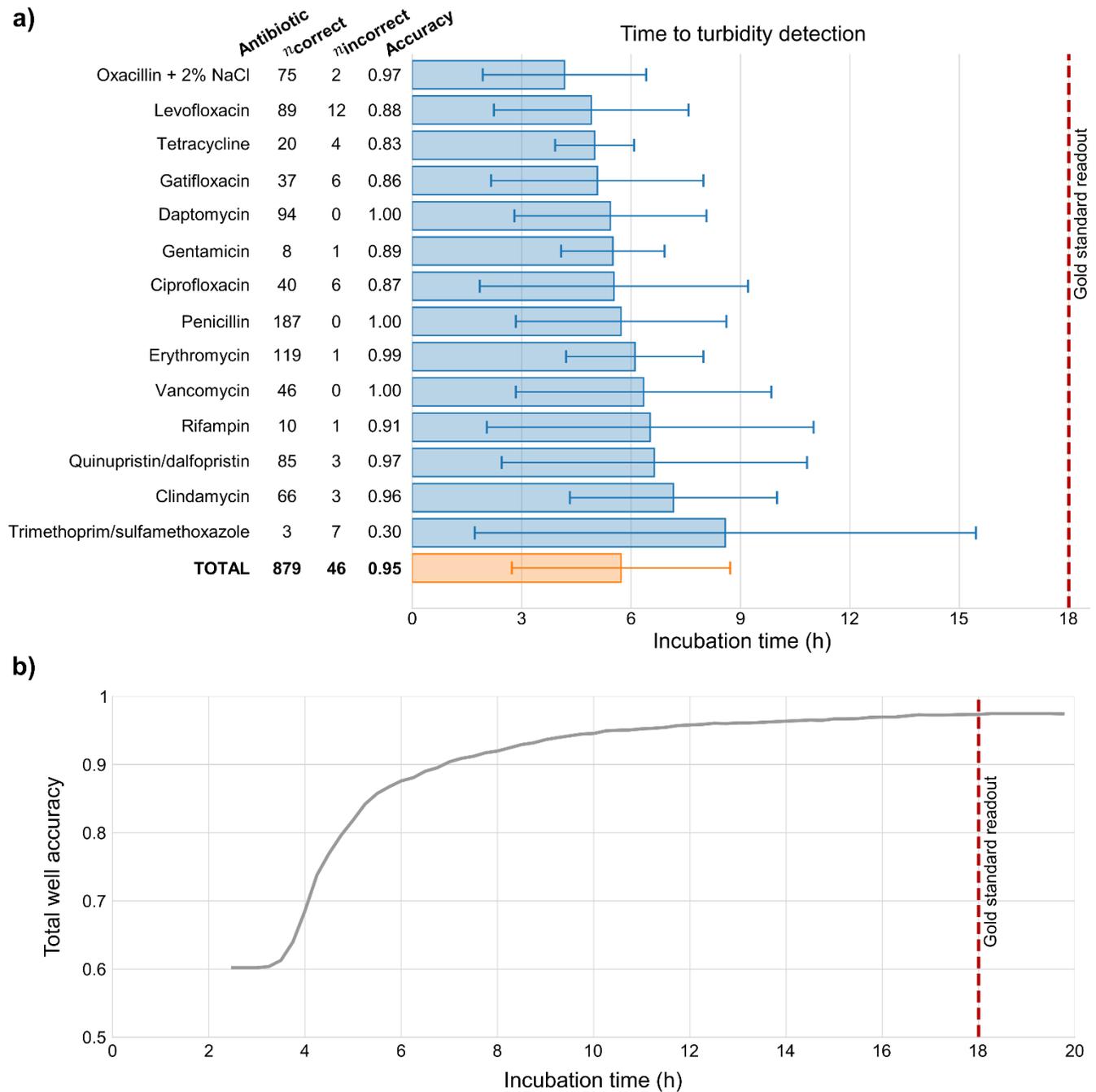

Figure 4. (a) Average time required for the panel of neural networks to make a correct turbidity prediction for each drug on blind testing isolates of *Staphylococcus aureus*. 95.03% of all turbid wells were correctly identified by the network, with the average turbid well requiring 5.72 h of incubation for automated detection. (b) Average well accuracy over the course of incubation. 90% of all wells were correctly classified after 7 h, and 95% after 10.5 h of incubation.

FDA limit (levofloxacin, ciprofloxacin, trimethoprim/sulfamethoxazole, and quinupristin/dalfopristin) each experienced only a single very major error. It was not possible to calculate vmj rate for vancomycin because no clinical isolates exhibited resistance to it. Using the data from Figure 5, the incubation times required to meet/surpass the FDA limits for EA/CA/maj/vmj rate are listed in Table 1. The system met the FDA-defined criteria for EA/CA for all 14 drugs after an average of 6.13 h and 6.98 h, respectively. The system met FDA criteria for major and very major error rates for 11 of 12 possible drugs after an average of 4.02 h, and 9 of 13 possible drugs after an average of 9.39 h, respectively. These results are in line with the performance on the validation data, demonstrating that the panel of networks is not overfit (Figures S5, S6, and Table S3).

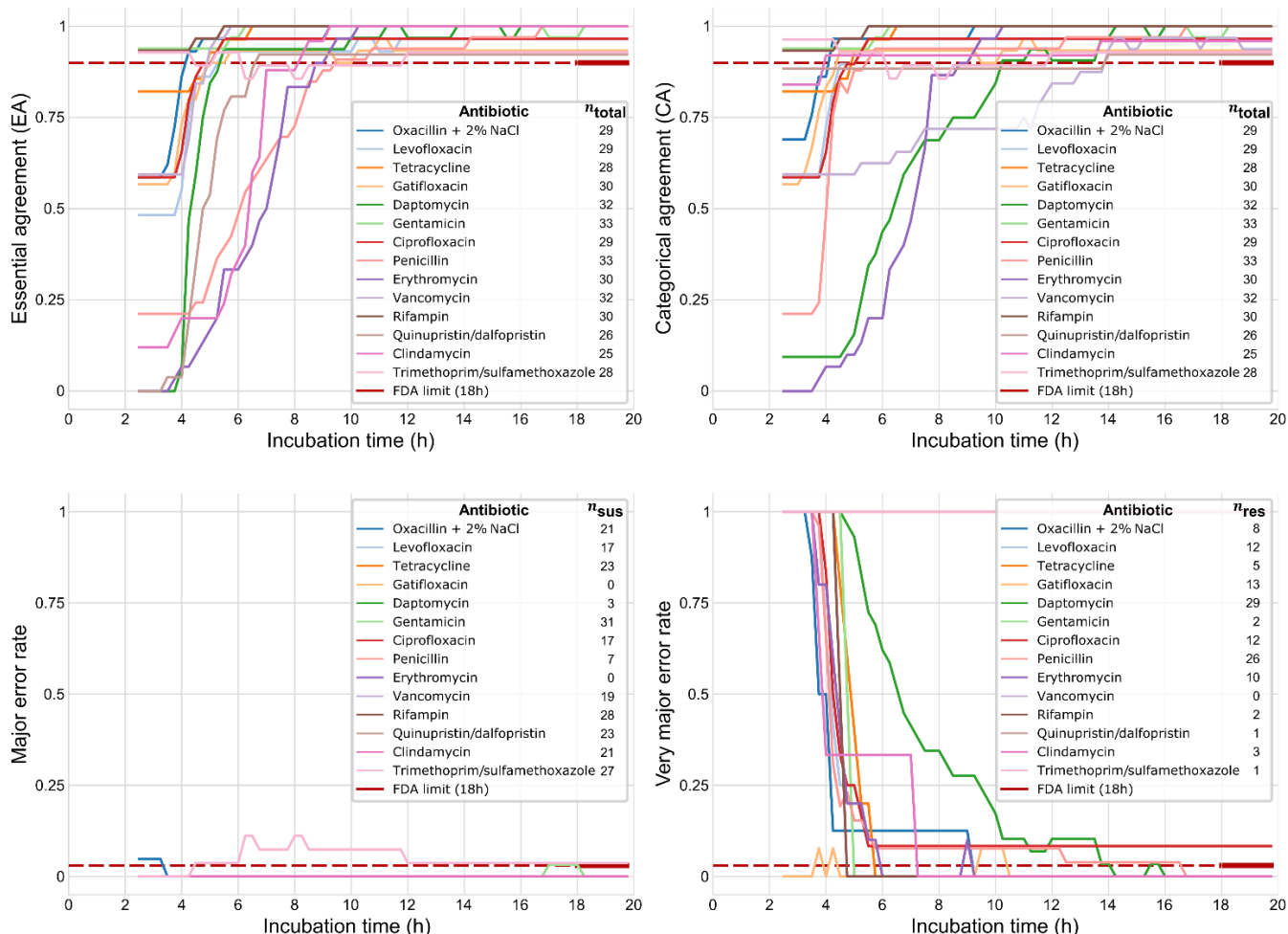

Figure 5. Essential agreement (EA), categorical agreement (CA), major error (maj) rate, and very major error (vmj) rate as a function of the incubation time for different antibiotics on blind testing isolates of *Staphylococcus aureus*. The second column in each plot legend indicates the number of samples for the corresponding metric (total number of valid samples for EA/CA, number of susceptible samples for maj rate, and number of resistant samples for vmj rate). EA and CA surpass the FDA limit of 90% for all 14 drugs well before the end of incubation. Maj rate remained below the FDA-permitted maximum of 3% for 11 of 12 possible drugs and vmj rate dropped below the FDA maximum of 3% before the end of incubation for 9 of 13 possible drugs.

The curve labeled "panel of networks" corresponds to the composite panel of neural networks, whereas "single network" refers to the best individual network (by validation loss) from the panel. Three additional panels of networks were also generated (from a sample size of 50 nine-fold cross-validations as before) using only images of a single illumination color (red, green, or blue). Finally, a logistic regression model was tested, as well as a simple threshold-based model, which classifies the well as turbid if at least two of the three most recent images have at least one fiber intensity lower than the threshold of 0.8876. This threshold value was determined by optimizing accuracy over the training patient plates.

The panel of networks showed the best performance across all four metrics, but there was only a slight penalty in EA and maj rate by using a single network. The networks that used only images of a single color fared considerably worse than the network or panel of networks using all three colors. Among the three colors we would not expect a large difference, but green did perform the best, possibly since it used twice the number of pixels per fiber due to the Bayer filter array on the CMOS image sensor. The logistic regression performed better than the single-color networks, but not as well as the three-color network or panel of networks. The threshold-based simple approach was the worst performer, with under 90% CA and over 10% vmj rate. Figures S1 and S2 show many examples of wells with predictions from the panel of networks, logistic regression, and threshold approach, demonstrating where the simpler models both failed to identify weak growth and falsely identified growth in non-turbid wells. Figures S7a,b show additional fiber intensities and network predictions for instances of a "skipped" well and wells with bubbles, respectively. These are well-known phenomena in AST and the network gave correct turbidity predictions in each case.

## Discussion

| Drug | Essential Agreement | Categorical Agreement | $n_{total}$ | Major error rate | $n_{susceptible}$ | Very major error rate | $n_{resistant}$ |
|---|---|---|---|---|---|---|---|
| Oxacillin + 2% NaCl | 4.25 | 4.25 | 29 | 3.5 | 21 | 9.25 | 8 |
| Levofloxacin | 5.5 | 5.25 | 29 | 2.5 | 17 | N/A | 12 |
| Tetracycline | 5 | 5 | 28 | 2.5 | 23 | 5.75 | 5 |
| Gatifloxacin | 5 | 4.5 | 30 | - | 0 | 10.5 | 13 |
| Daptomycin | 5.5 | 10.25 | 32 | 2.5 | 3 | 16 | 29 |
| Gentamicin | 2.5 | 2.5 | 33 | 18.25 | 31 | 5 | 2 |
| Ciprofloxacin | 5.25 | 5.25 | 29 | 2.5 | 17 | N/A | 12 |
| Penicillin | 9.5 | 5.5 | 33 | 2.5 | 7 | 16.75 | 26 |
| Erythromycin | 8.75 | 8.75 | 30 | - | 0 | 9.25 | 10 |
| Vancomycin | 5 | 14 | 32 | 2.5 | 19 | - | 0 |
| Rifampin | 2.5 | 2.5 | 30 | 2.5 | 28 | 4.75 | 2 |
| Quinupristin/ dalfopristin | 6.75 | 14 | 26 | 2.5 | 23 | N/A | 1 |
| Clindamycin | 8.25 | 4 | 25 | 2.5 | 21 | 7.25 | 3 |
| Trimethoprim/ sulfamethoxazole | 12 | 12 | 28 | N/A | 27 | N/A | 1 |
| **AVERAGE** | **6.13** | **6.98** | | **4.02** | | **9.39** | |

Table 1. Incubation time (h) required to meet FDA criteria by drug for blind testing isolates of *S. aureus*.

Our system demonstrates the ability to detect turbidity and quantify resistance much sooner than the gold standard method, which requires at least 18–24 h. On the blind testing data, the system made only one major error and four very major errors across all 33 patients with 14 drugs each. The fiber intensities and network predictions for the wells corresponding to each of these errors are shown in Figures S8–10 along with an image of the wells captured with a smartphone camera at the end of incubation. Each drug for which our system exceeded the FDA-defined limit of 3% for maj or vmj rate only experienced a single error. With additional testing samples, the maj/vmj rates may drop below 3% for all drugs. In addition, the FDA defines major and very major errors as misclassification of a susceptible/resistant organism as resistant/susceptible. However, the one major error and one of the four very major errors from the network's predictions did not include a predicted susceptibility because the predicted MIC was undefined (known as a "skipped well"). We report these instances as major/very major errors, but they can be thought of as inconclusive results, for which a human could be notified to read the MIC manually or decide to repeat the test.

The performance of the system demonstrates the potential to enable automated, cost-effective susceptibility testing with early results in resource-limited laboratories. Unlike the gold standard BMD method, our system does not require a full 18–24 h incubation or a trained technologist for plate readout and, unlike microscopy-based solutions, it requires no mechanical scanning components or bulky, costly hardware. Cost and access to trained personnel are primary factors that currently limit the reach of AST in developing regions. Our system also uses standard 96-well plates, which would allow it to more rapidly integrate with typical clinical workflow. In addition, because the system autonomously captures images during incubation without the need to remove the plate from the incubator, early results for drugs showing strong resistance can be sent to the physician as soon as they are available, while the device continues to monitor growth in the wells with the remaining drugs. Due to the phenotypic nature of our sensing mechanism, we believe it can be extended to almost any type of bacteria, or other plate-based tasks such as enzyme-linked immunosorbent assays (ELISA) and culture samples. The fiber-based subsampling of the wells could enable the streamlining of daily laboratory tasks with robust, automated readout in a compact form factor.

From Figure 6, it is clear that the panel of neural networks gave the best performance on the blind testing data, demonstrating an ability to discern nuanced patterns in fiber intensities. While a desktop computer was used to train the panel of networks, due to the rapidly decreasing cost of computation in embedded systems, future training could be performed on the Raspberry Pi or other compact device. In addition, a single network showed only a modest drop in performance, which could shorten computation time. Because the networks were not given knowledge of the well, drug, or concentration when making predictions, they also learned a model of turbidity that is quite general, instead of overfitting to the specifics of the plate or drugs used in the experiments. Because the second technologist who made ground truth readings for the testing data was not used for the training/validation data, the system demonstrated an ability to generalize beyond the specific patterns of an individual human reader.

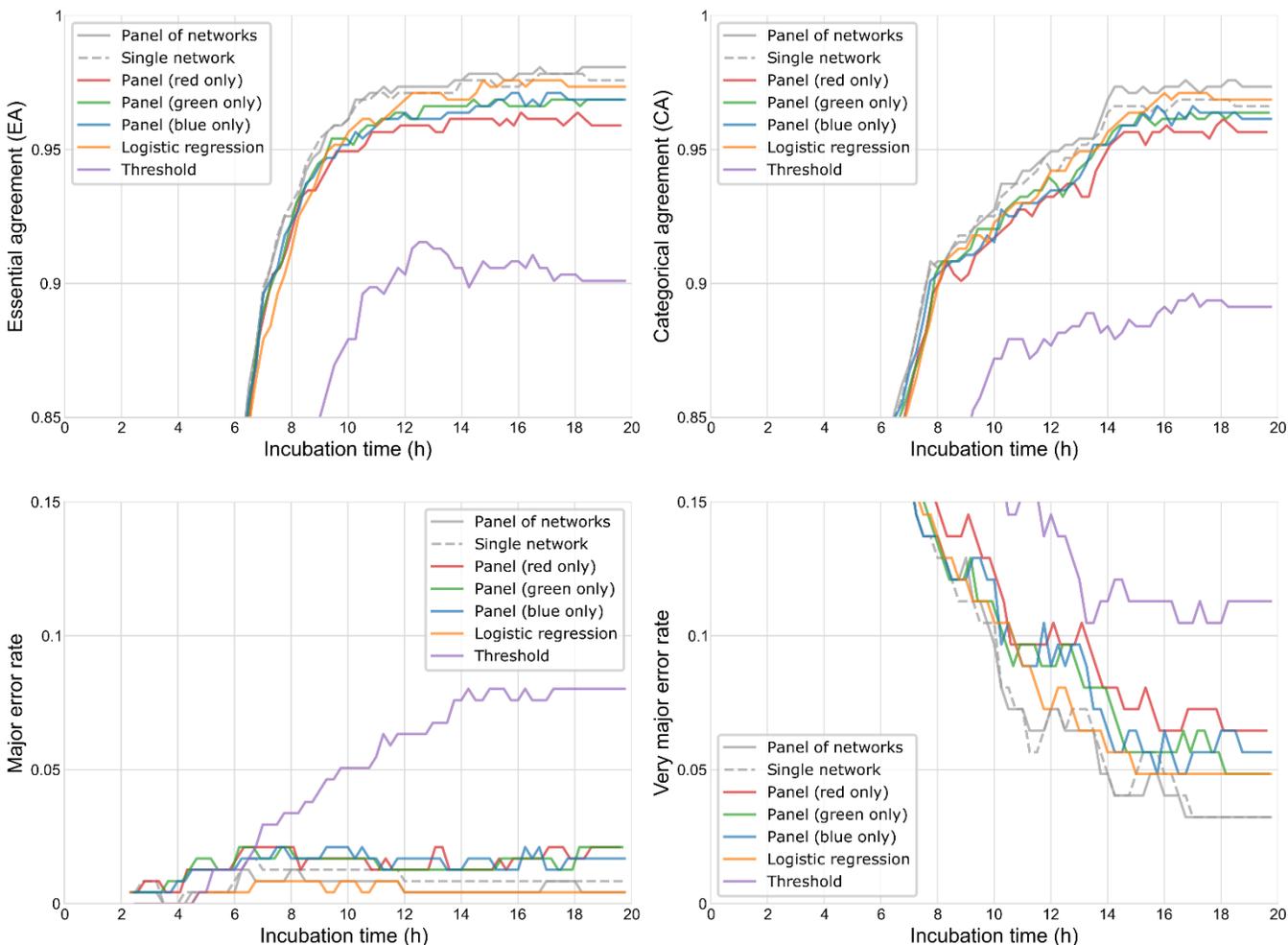

Figure 6. Essential agreement (EA), categorical agreement (CA), major error (maj) rate, and very major error (vmj) rate averaged over all the drugs for various models for blind testing data. "Panel of networks" refers to the panel of nine neural networks trained via cross-validation. "Single network" is only the best of the nine networks (by validation loss). "Panel (red/green/blue)" is a panel of networks that only uses images of the specified color. "Threshold" is a simple threshold-based approach in which a well is classified as turbid if fiber intensities fall below a specified threshold (see Methods section).

## Conclusion

The presented system demonstrates the ability to conduct AST much faster than the gold standard method of incubation for 18–24 h followed by visual inspection. The time savings is critical to ensuring patients receive the most effective, targeted antibiotics and to limit the global rise in antimicrobial resistance. Our system also removes the need for a trained medical technologist and integrates with the standard clinical workflow using an incubator and 96-well microplates. The system is cost-effective due to the use of off-the-shelf components and could be particularly suited to resource-limited laboratories in developing regions, where antimicrobial resistance is predicted to cause the most deaths and access to trained personnel is limited.

## Methods

### Imaging System

The AST system illumination is composed of two 8x8 arrays of individually addressable RGB LEDs (Adafruit Industries) whose pulse width modulation brightness is set by 2 Trinket microcontrollers (Adafruit Industries). The system contains 2016 0.75 mm plastic optical fibers (CK-30, Industrial Fiber Optics Inc.), which were epoxied and polished with a handheld polishing tool. The two common fiber bundles each contain fibers for half of the 96-well plate. They are each imaged by the combination of a 10.0 mm diameter × 50.0 mm focal length plano-convex lens (Edmund Optics) and a Raspberry Pi Camera Module V2 (Newark) with 1.12 μm × 1.12 μm pixel size. The two cameras are controlled by two Raspberry Pi 3 Model B computers. Images are captured in raw 10-bit format at 8.1 MP.

**Image Processing and Neural Network**

Image processing was performed in MATLAB (MathWorks) and neural network training/testing was performed in Python using TensorFlow 1.14 (Google). The logistic regression model was created in Python with the scikit-learn library (David Cournapeau), using the saga solver, the elasticnet penalty with an L1 ratio of 0.9, and an inverse regularization strength of C = 2.

**Clinical Testing**

All experiments were performed at the UCLA Clinical Microbiology Laboratory. The AST system was placed inside a 2-cubic foot incubator (Binder) for the duration of the experiments. Initial experiments were performed using the ATCC43300 strain of MRSA in 47 plates. Confirmed clinical *S. aureus* isolates collected at the UCLA Clinical Microbiology Laboratory were tested on the platform for the remainder of the experiments. *S. aureus* isolates were prepared to a 0.5 McFarland standard in sterile water and 50 μL of this suspension was transferred into 11 mL of Mueller Hinton Broth. The dilution was inoculated into 96-well microplates (100 μL of bacterial suspension per well) containing a commercially available Gram-positive antibiotic panel (Sensititre Gram Positive MIC plates, ThermoFisher Scientific) shown in Table S4. Following bacterial inoculation, single plates were loaded into the incubator for 18–19 h. At the end of incubation, plates were removed and turbidity was manually assessed by trained personnel. For training/validation data (51 clinical plates), plates were read by a single reader. For testing data (33 clinical plates), plates were read by two readers to assess and mitigate interpersonal variances among readers. MIC was determined by identification of the first well without turbidity for increasing drug concentrations. Interpretation of susceptibility was determined in accordance to Clinical & Laboratory Institute Standards 2019.[8]

## AUTHOR INFORMATION

### Corresponding Author

* ozcan@ucla.edu

### Funding Sources

The Ozcan Research Group at UCLA acknowledges the support of Kairos Ventures, Technology Development Group at UCLA, National Science Foundation (NSF) Engineering Research Center (ERC, PATHS-UP), and the Howard Hughes Medical Institute. C. B. acknowledges the support of the NSF GRFP fellowship.

## ACKNOWLEDGMENT

The authors acknowledge the UCLA Clinical Microbiology Laboratory staff for their guidance and expertise.